\documentclass[twocolumn,english,prb,showpacs,superscriptaddress]{revtex4-1}
\usepackage[colorlinks=true,urlcolor=blue,citecolor=blue,linkcolor=blue]{hyperref}

\usepackage[T1]{fontenc}
\usepackage[latin9]{inputenc}
\usepackage{amssymb}
\usepackage{graphicx}
\usepackage{amsmath,color}
\usepackage{mathrsfs}
\usepackage{float}
\usepackage{indentfirst}
\usepackage{subfigure}
\usepackage{multirow}
\usepackage{tabu}
\usepackage{booktabs}
\usepackage{txfonts}
		
\usepackage[euler]{textgreek}

\usepackage{graphicx}
\usepackage{bm}

\newcommand{\Eq}[1]{Eq.~(\ref{#1})}

\begin{document}

\preprint{APS/123-QED}

\title{Universal boundary entropies in conformal field theory: \\ A quantum Monte Carlo study}%

\author{Wei Tang}
 \affiliation{International Center for Quantum Materials, School of Physics, Peking University, Beijing 100871, China}

\author{Lei Chen}
\author{Wei Li}
 \affiliation{Department of Physics, Key Laboratory of Micro-Nano Measurement-Manipulation and Physics
(Ministry of Education), Beihang University, Beijing 100191, China}

\author{X. C. Xie}
 \affiliation{International Center for Quantum Materials, School of Physics, Peking University, Beijing 100871, China}

\author{Hong-Hao Tu}
 \email{h.tu@lmu.de}
\affiliation{Physics Department, Arnold Sommerfeld Center for Theoretical Physics and Center for NanoScience,
Ludwig-Maximilians-Universit\"at M\"unchen, 80333 M\"unchen, Germany}

\author{Lei Wang}
 \email{wanglei@iphy.ac.cn}
 \affiliation{Beijing National Lab for Condensed Matter Physics and Institute of Physics, Chinese Academy of Sciences, Beijing 100190, China
}

\date{\today}

\begin{abstract}
  Recently, entropy corrections on nonorientable manifolds such as the Klein bottle are proposed as a universal characterization of critical systems with an emergent conformal field theory (CFT). We show that entropy correction on the Klein bottle can be interpreted as a boundary effect via transforming the Klein bottle into an orientable manifold with nonlocal boundary interactions. The interpretation reveals the conceptual connection of the Klein bottle entropy with the celebrated Affleck-Ludwig entropy in boundary CFT. We propose a generic scheme to extract these universal boundary entropies from quantum Monte Carlo calculation of partition function ratios in lattice models. Our numerical results on the Affleck-Ludwig entropy and Klein bottle entropy for the $q$-state quantum Potts chains with $q=2,3$ show excellent agreement with the CFT predictions. For the quantum Potts chain with $q=4$, the Klein bottle entropy slightly deviates from the CFT prediction, which is possibly due to marginally irrelevant terms in the low-energy effective theory.
\end{abstract}

\maketitle

\section{Introduction}

Critical points of continuous phase transitions are described by the
renormalization group fixed points with divergent correlation
length{\cite{altland_condensed_2010}}. For one-dimensional (1D) quantum systems,
or equivalently, two-dimensional classical systems, these fixed points can be
classified by conformal field theory
(CFT){\cite{francesco_conformal_2012,fradkin_field_2013}}, due to the
conformal invariance at these critical points. Moreover, boundary CFT {\cite{cardy_conformal_1984,cardy_boundary_2004}} offers a powerful description for critical systems with boundaries. Affleck-Ludwig (AL) entropy is a universal boundary entropy appearing in the
boundary CFT{\cite{affleck_universal_1991}} that originates from the open boundary of
the path-integral manifold and depends on the universality class of the
boundary conditions. AL entropy is shown to be nonincreasing under the
boundary renormalization group
flow{\cite{affleck_universal_1991,friedan_boundary_2004}} and thus determines
the relative stability of various phases. AL entropy is also closely related to Kondo
problems{\cite{affleck_conformal_1995,affleck_entanglement_2009,altland_multichannel_2014}}, quantum point contacts{\cite{teo_critical_2009}}, and
the entanglement
entropy{\cite{calabrese_entanglement_2004,laflorencie_boundary_2006,zhou_entanglement_2006,calabrese_entanglement_2009,fradkin_scaling_2009,hsu_universal_2009,kitaev_topological_2006,levin_detecting_2006,fendley_topological_2007}}. Despite its
importance, unbiased numerical determination of AL entropy has been a challenging
task, where most existing calculations exploit its relation to the entanglement properties of ground state wavefunctions {\cite{zhou_entanglement_2006,shi_finite-size_2010,stephan_geometric_2010,hu_geometric_2011,liu_geometric_2012}}.

Recently, Ref.~{\onlinecite{tu_characterizing_2017}} showed that another universal entropy emerges for CFT defined on the Klein bottles {\cite{tu_characterizing_2017}}. The Klein bottle entropy originates from the nonorientablity of the Klein bottle manifold. Besides being used to characterize the CFT, the Klein bottle entropy can also be used to accurately pin point quantum critical points, even those without local order parameters{\cite{chen_universal_2017}}.

In this paper, we reveal conceptual connections of the Klein bottle entropy and the boundary AL entropy. Although the former is defined on Klein bottle which has no boundary, we show that after a cutting-and-sewing transformation of the Klein bottle manifold, the Klein bottle entropy can be attributed to nonlocal interactions emerging at the manifold boundary. Revealing such connection provides a unified way to compute the AL entropy and the Klein bottle entropy in numerical calculations. We present an efficient method to extract these universal boundary entropies of CFT from quantum Monte Carlo (QMC) simulation of lattice models. Extended ensemble QMC simulation allows us to extract universal boundary entropies directly from thermodynamic quantities such as free energy difference on various manifold. As an application of the method, we compute the AL entropy and the Klein bottle entropy of $q$-state quantum Potts chains at their quantum critical points.

This paper is organized as follows. In Sec.~\ref{ratio}, we briefly review
the CFT predictions for the partition functions on various path-integral
manifolds and then show that the universal boundary entropies can be extracted
as partition function ratios. The resemblance of the AL entropy and the Klein
entropy is discussed. In Sec.~\ref{ModelAndMethod}, we introduce the $q$-state quantum
Potts Hamiltonian, the boundary entropies of which are computed. We then introduce the extended
ensemble QMC method for the partition function ratios.
In Sec.~\ref{result}, we present the numerical results and compare them with CFT predictions.
Sec.~\ref{conclusion} summarizes the results and provides an outlook for future directions. We present details of our quantum Monte Carlo implementation in Appendix \ref{app-update-scheme} and summarize various conformal boundary conditions and their treatment in QMC in Appendix \ref{appendix:complete-set-of-bc} and Appendix \ref{appendix-reweighting}.

\begin{figure}[!htb]
  {\includegraphics[width=\columnwidth]{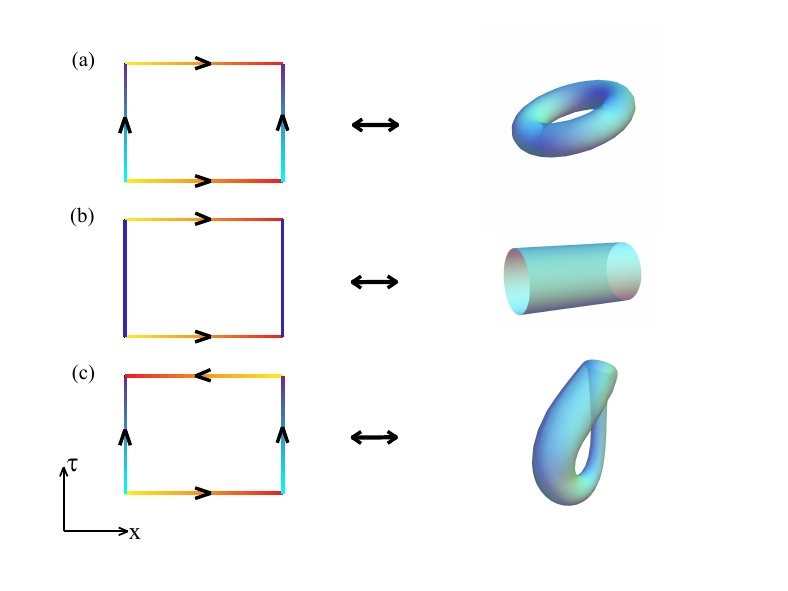}}
  \caption{ (a) Path integral manifold of a 1D quantum system with periodic boundary condition in the spatial direction is an equivalent to a torus. The direction of black arrows indicate how the space or time boundaries are joined. (b) With open boundary condition in the spatial direction, the
  path integral manifold is a cylinder. (c) The path integral manifold of Eq.~(\ref{eq:ZK}) corresponds to a Klein bottle due to the spatial reflection inserted in the trace.
  \label{fig:topology}}
\end{figure}

\section{Universal boundary entropies on various manifolds}\label{ratio}

In the path integral formulation, a 1D quantum system can be formulated as a $(1+1)$-dimensional classical system. The additional dimension corresponds to the imaginary time direction. With the periodic boundary condition, the path integral manifold is a
torus [see Fig.~\ref{fig:topology}(a)]. For $(1+1)$-dimensional critical systems with large $\beta$ and the system size $L \gg
v\beta$, the logarithm of the partition
function takes the following form{\cite{affleck_universal_1986,blote_conformal_1986}}

\begin{equation}
  \mathrm{ln} Z^{\mathcal{T}} (L, \beta) = - f_0 \beta L + \frac{\pi c}{6
  \beta v} L + \mathcal{O} \left( \frac{1}{\beta^2} \right), \label{eq:z-pbc}
\end{equation}
where $f_0$ is the nonuniversal free energy density, $c$ is the central
charge, and $v$ is the speed of ``light'' of the CFT. Notably, the leading order of the
correction term gives a bulk entropy $\frac{\pi c}{6 \beta v} L$, which is
proportional to the central charge $c$. 

For the same system with open boundary
condition, whose path integral manifold is a cylinder with open boundaries [see Fig.~\ref{fig:topology}(b)], there are additional corrections to the free energy
\begin{equation}
  \mathrm{ln} Z^{\mathcal{C}} (L, \beta) = - f_0 \beta L + \frac{\pi c}{6
  \beta v} L + S_{\mathrm{AL}}   - f_\mathrm{b} \beta + \mathcal{O} \left( \frac{1}{\beta^2}
  \right) . \label{eq:z-obc}
\end{equation}
Here $S_{\mathrm{AL}}$ is the AL entropy{\cite{affleck_universal_1991}}, which is universal and only depends on the CFT and the boundary conditions\footnote{In order to obtain the universal AL entropy, here the boundary conditions should be chosen to be conformal boundary conditions}. In addition, there is a nonuniversal term $- f_\mathrm{b} \beta$ in
(\ref{eq:z-obc}), which originates from the surface free energy on the open
boundaries of the cylinder. Comparing Eqs.~(\ref{eq:z-pbc}) and (\ref{eq:z-obc}), one can obtain the boundary corrections from the partition function ratio
\begin{equation}
  \mathrm{ln}\left[ \frac{Z^{\mathcal{C}} (L, \beta)}{Z^{\mathcal{T}} (L, \beta)} \right]=
  S_{\mathrm{AL}} - f_\mathrm{b} \beta + \mathcal{O} \left( \frac{1}{\beta^2} \right) .
  \label{eq:SAL}
\end{equation}
A linear extrapolation of the partition function ratio \Eq{eq:SAL} will give universal AL entropy as the intercept and the surface free energy as the slope.

Next, consider a partition function of a periodic chain with an inserted spatial reflection operator
\begin{equation}
Z^{\mathcal{K}}= \mathrm{Tr}
\left(\hat{P} e^{- \beta \hat{H}}\right),
\label{eq:ZK}
\end{equation}
 where $\hat{P}$ swap the state on site $i$  and site $L - i + 1$ of the 1D chain. As a result,
 $Z^{\mathcal{K}} = \sum_{\ensuremath{\boldsymbol{\sigma}} } \langle \ensuremath{\boldsymbol{\sigma}} |e^{-\beta \hat{H}} |\overleftarrow{ \ensuremath{\boldsymbol{\sigma}}}\rangle$, where $| \ensuremath{\boldsymbol{\sigma}} \rangle \equiv | \sigma_1
\rangle \otimes | \sigma_2 \rangle \otimes \ldots \otimes | \sigma_L \rangle$ is the basis state, and  $| \overleftarrow{\ensuremath{\boldsymbol{\sigma}}} \rangle \equiv | \sigma_L
\rangle \otimes | \sigma_{L-1} \rangle \otimes \ldots \otimes | \sigma_1 \rangle$ is the spatial reflected basis state. In the path integral formulation, the worldlines of the partition function \Eq{eq:ZK} twist before joining in the imaginary time direction. The corresponding path-integral manifold is therefore topologically equivalent to a Klein bottle [see Fig.~\ref{fig:topology}(c)]. In this case, a universal entropy emerges in the free energy provided the reflection operator switches the left and right movers in CFT~\cite{tu_characterizing_2017}
\begin{equation}
  \mathrm{ln} Z^{\mathcal{K}} (L, \beta) = - f_0 \beta L + \frac{\pi c}{24
  \beta v} L + S_{\mathrm{KB}} + \mathcal{O} \left( \frac{1}{\beta^2} \right) .
  \label{klein-bottle-partition-function}
\end{equation}
Here $S_{\mathrm{KB}} = \mathrm{ln} \left(\sum_a M_{a a} d_a / \mathcal{D}\right)$ is the Klein bottle entropy that originates
from the nonorientablity of the manifold,
where $d_a$'s are the quantum dimensions of the primary fields of the CFT and $\mathcal{D} =
\sqrt{\sum_a d_a^2}$ is the total quantum
dimension{\cite{tu_characterizing_2017}}. For diagonal CFT partition
functions, $M_{a a} = 1$ for $\forall a$.

\begin{figure}[!htb]
 {\includegraphics[width=\columnwidth]{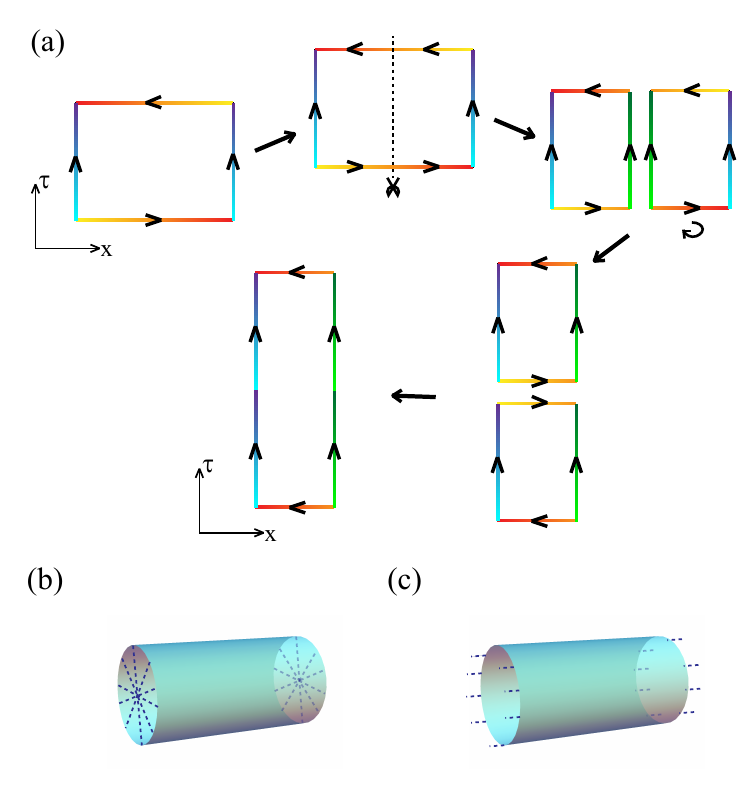}}
  \caption{ (a) Transforming a Klein bottle path integral manifold into an orientable manifold via cutting, flipping, and sewing the manifolds. 
  (b) A Klein bottle is transformed into a cylinder with nonlocal interactions along
  the time direction as indicated
  by the dashed lines on the cylinder boundaries. (c) A torus corresponds to 
  a cylinder with two spatial boundaries joined by the nonlocal interactions indicated by the dashed lines.
  \label{klein-twist-fig}}
\end{figure}

Remarkably, the bulk
entropy $\frac{\pi c}{24 \beta v} L$ in the Klein bottle manifold \Eq{klein-bottle-partition-function} is distinct from the bulk
entropy $\frac{\pi c}{6 \beta v} L$ of the torus \Eq{eq:z-pbc}. This can
be understood by the transformation{\cite{blumenhagen_introduction_2009}} illustrated in Fig.~\ref{klein-twist-fig}(a). 
By cutting along the imaginary-time direction of the Klein bottle, flipping one piece and sewing it back to another piece along the spatial direction, one obtains a manifold with doubled inverse temperature in the time direction and halved length in the spatial direction. When rolling the resulting manifold into a cylinder it is clear that there are nonlocal interactions along the time direction on the two spatial boundaries, as shown in Fig.~\ref{klein-twist-fig}(b). In comparison, a torus manifold can be viewed as a cylinder with nonlocal interactions along the spatial direction which joins the two cylindrical boundaries [see Fig.~\ref{klein-twist-fig}(c)]. In this regard, the universal entropy $S_{\mathrm{KB}}$ is a ``boundary effect'', which bears strong resemblance to the AL entropy. As a result of Eqs.~(\ref{eq:z-pbc}) and (\ref{eq:ZK}), and in accordance to the pictorial considerations in Fig.~\ref{klein-twist-fig}, the Klein bottle entropy can be extracted from the partition function ratio
\begin{equation}
  \mathrm{ln} \left [\frac{Z^{\mathcal{K}} (2 L, \beta / 2)}{Z^{\mathcal{T}} (L,
  \beta)}\right] = S_{\mathrm{KB}} + \mathcal{O} \left( \frac{1}{\beta^2} \right) . \label{SKB}
\end{equation}
Compared to \Eq{eq:SAL}, calculating $S_{\mathrm{KB}}$ from the partition function ratio is free of nonuniversal surface free energy
term because the Klein bottle manifold has no boundaries. \footnote{A caveat of the calculation is that the lattice reflection operator $\hat{P}$ may not play exactly the same role as the CFT~\cite{tu_characterizing_2017}, thus the computed ratio may deviate from the CFT prediction.}

\section{Model and Methods}\label{ModelAndMethod}

\subsection{\boldmath{$q$}-state quantum Potts chain}

We consider the Klein bottle entropy and the Affleck-Ludwig entropy in the
$q$-state quantum Potts chain. The Hamiltonian reads {\cite{solyom_renormalization-group_1981}}
\begin{equation}
  \hat{H} = - J \sum_{\langle i, j\rangle} \sum_{k = 1}^{q - 1} \hat{\sigma}_i^k
  \hat{\sigma}_{j}^{q - k} - \Gamma \sum_{i = 1}^L \sum_{k = 1}^{q - 1}
  \hat{\tau}_i^k, \label{q-potts-hamiltonian}
\end{equation}
where
\begin{equation}
  \hat{\sigma}_{i} = \left( \begin{array}{ccccc}
    1 &  &  &  & \\
    & \omega &  &  & \\
    &  & \omega^2 &  & \\
    &  &  & \ddots & \\
    &  &  &  & \omega^{q - 1}
  \end{array} \right), \hat{\tau}_{i} = \left( \begin{array}{ccccc}
    0 & 1 &  &  & \\
    & 0 & 1 &  & \\
    &  & 0 & \ddots & \\
    &  &  & \ddots & 1\\
    1 &  &  &  & 0
  \end{array} \right),
\end{equation}
and $\omega = \exp (2 \pi \mathrm{i} / q)$. The first term represents the Potts coupling between the neighboring sites, and the second
part is an analog of the transverse field. The critical point of this
model{\cite{solyom_renormalization-group_1981}} is $J = \Gamma$. In the case
of ferromagnetic coupling ($J > 0$), at zero temperature, when $J > \Gamma$,
the system is in the ordered phase, while for $J < \Gamma$, the system is in
the quantum disordered phase.

Recently, Ref.~{\onlinecite{ding_monte_2017}} studied the quantum Potts model using the stochastic series expansion quantum Monte Carlo method. In our work, we implement generic continuous-time path integral QMC code for the $q$-state quantum Potts model. Simulation of the model is unbiased on a finite-size lattice. Details of the QMC implementation are included in
Appendix \ref{app-update-scheme}.

\subsection{Extended ensemble Monte Carlo Method for partition function
ratios}
We compute the ratio of two partition functions using an extended ensemble simulation. This approach avoids computing the two
partition functions separately. 

The partition function of the extended ensemble is a summation of two partition functions
\begin{equation}
  Z = Z^{\mathcal{C}} + Z^{\mathcal{T}}= \sum_{\eta \in \{\mathcal{C}, \mathcal{T}\} } \sum _{ \mathscr{C}} w^{\eta}(\mathscr{C}) \, ,
  \label{eq:EE}
\end{equation}
where in the second equality we combine the sum over ensemble label $\eta$ and the Monte Carlo configuration $\mathscr{C}$.  $w^{\eta}(\mathscr{C})$ is the Boltzmann weight of the Monte Carlo simulation. The simulation treats the update of the QMC configuration $\mathscr{C}$ and the ensemble label $\eta$ on the equal footing. Thus, there is also an update of switching the label. Appendix \ref{app-update-scheme} contains details about the Monte Carlo simulations.


To obtain the ratio of $Z^{\mathcal{C}}$ and $Z^{\mathcal{T}}$,
%
we use the Bennett acceptance ratio {\cite{bennett_efficient_1976}} method, which was originally developed for estimating
the free energy difference between two ensembles. The estimator of partition function ratio is
\begin{equation}
  \frac{Z^{\mathcal{C}}}{Z^{\mathcal{T}}} = \frac{\left\langle \left( 1 +
  w^{\mathcal{T}} / w^{\mathcal{C}} \right)^{- 1} \right\rangle}{\left\langle
  \left( 1 + w^{\mathcal{C}} / w^{\mathcal{T}} \right)^{- 1}\right \rangle} \, , \label{eq:BAR}
\end{equation}
where the expectation value $\left\langle \cdot \right\rangle $ refers to the average sampled in the extended ensemble Eq.~(\ref{eq:EE}).

For the ratio in \Eq{SKB} we devised an extended ensemble similar to \Eq{eq:EE}.  We represent the configurations in the Klein bottle ensemble using the transformation of Fig.~\ref{klein-twist-fig}(a),
so that the two ensembles $Z^{\mathcal{K}} (2L, \beta / 2)$  and $Z^{\mathcal{T}} (L, \beta)$ only differ by the boundary interactions shown in Figs.~\ref{klein-twist-fig}(b) and \ref{klein-twist-fig}(c).

\section{Results}\label{result}

\subsection{Affleck-Ludwig boundary entropy}

The AL entropy depends on the CFT of the system and the boundary
conditions. Among all boundary conditions, the conformal boundary
conditions play an important
role{\cite{cardy_boundary_1989,affleck_conformal_1995,cardy_boundary_2004}}.
For the $q$-state quantum Potts chain ($q = 2, 3$) in which we calculate the AL
entropy, the complete set of conformal boundary conditions has been obtained
by the boundary CFT{\cite{cardy_boundary_1989,saleur_relations_1989,affleck_boundary_1998,fuchs_completeness_1998}}. We summarize these conformal boundary conditions and their lattice realizations in Appendix
\ref{appendix:complete-set-of-bc}.

It is worth mentioning that, to achieve certain boundary conditions,
one needs to add a large pinning field on the boundary sites of the quantum Potts chain. The pinning
field term will enlarge the cylinder partition function by a 
large factor, so we have to perform a reweighting procedure during the QMC
simulation, the details of which are discussed in Appendix
\ref{appendix-reweighting}.

\begin{figure}[!t]
  {\includegraphics[width=\columnwidth]{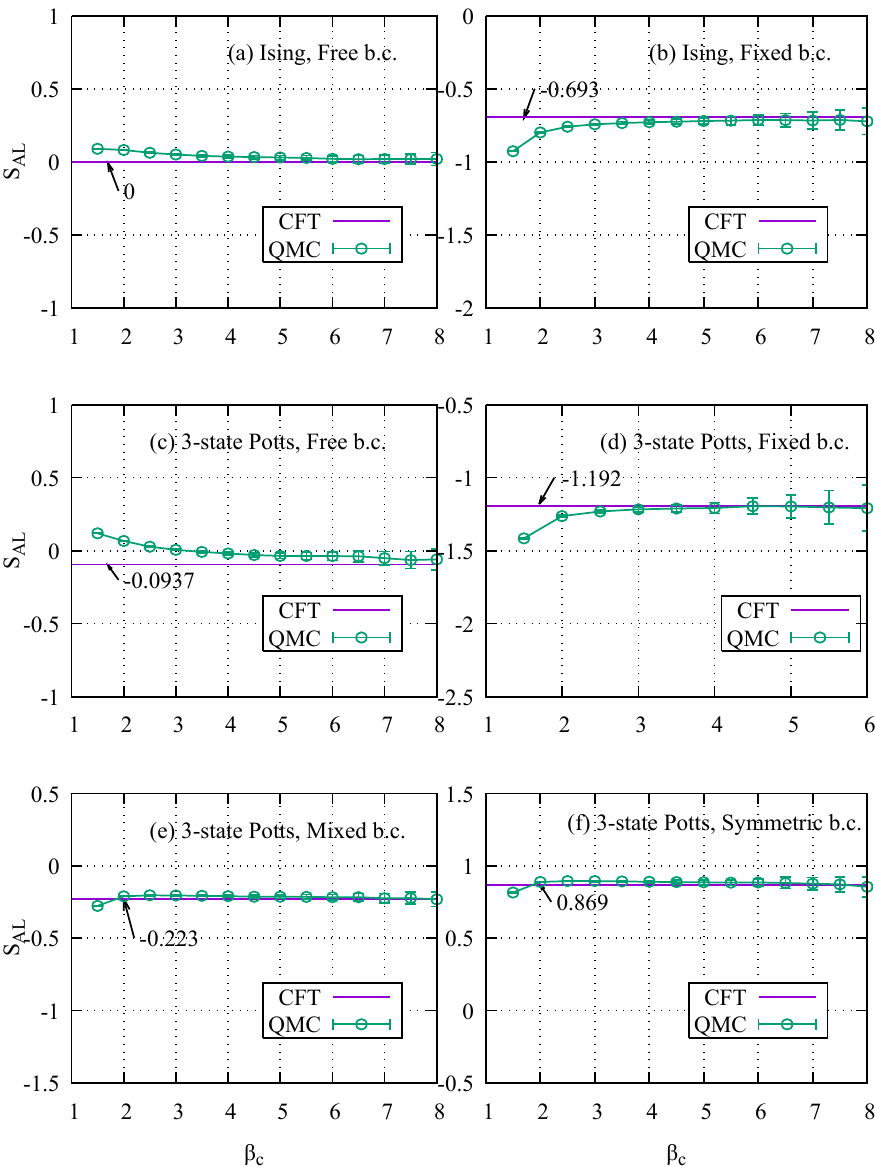}}
  \caption{Results of the AL entropy extracted from a linear fit of the QMC result of \Eq{eq:SAL}, including Ising
  model with (a) free and (b) fixed boundary conditions and three-state
  Potts model with (c) free, (d) fixed, (e) mixed, and (f) symmetric boundary conditions (abbreviated as b.c. in the figures).
  The intercept of the linear regression over
  $[\beta_c - 1, \beta_c + 1]$
  is plotted versus the center of the fitting range  $\beta_c$. The horizontal lines indicate the CFT predictions of the AL entropies. The size of
  the systems is $L = 500$.
  \label{fig:al-entropy-result}
 }
\end{figure}

The AL entropy is calculated in a $q$-state quantum Potts chain for $q = 2, 3$ with the
corresponding conformal boundary conditions. Throughout the calculation we consider the same
boundary conditions at left and right edges. According to Eq.~(\ref{eq:SAL}), to obtain the AL
entropy $S_{\mathrm{AL}}$, we calculate $\mathrm{ln} \left( Z^{\mathcal{C}} /
Z^{\mathcal{T}}\right)$ for different temperatures with the system size
fixed, and then perform a linear extrapolation with $\beta$. Due to the
existence of the correction term $\mathcal{O} (1 / \beta^2)$, in order to obtain the
correct value of $S_{\mathrm{AL}}$, the linear regression should be performed
at large enough $\beta$, but also with the condition $ L\gg v\beta$ satisfied.
In our simulation, the system size is chosen as $L = 500$, and we perform the
linear regression over the range $[\beta_c - 1, \beta_c + 1]$, with $\beta_c$
gradually growing. We plot the intercept of the linear regression, which is
an estimate of $S_{\mathrm{AL}}$, with respect to the value of $\beta_c$, the
center of the fitting range, as shown in Fig.~\ref{fig:al-entropy-result}. The
CFT results of the AL entropy of different type of
systems{\cite{affleck_boundary_1998}} and boundary conditions are also marked
as a horizontal line in the figure correspondingly. It can be seen that the
numerical results and the CFT predictions are consistent with each other, except
the deviation at small $\beta$ which originates from the $\mathcal{O} (1 / \beta^2)$
correction. Furthermore, from the slope of the linear extrapolation one also obtains the value of the surface free energy. For example, for the free boundary condition of the three-state quantum Potts chain, the numerical result of the surface free energy is $f_\mathrm{b} \approx 0.5988(8)$ (fitting range $\beta=6\sim 9$) which coincides with the Bethe ansatz result\cite{batchelor_surface_1990, alcaraz_conformal_1988} $f_\mathrm{b} =  3\sqrt{3}/2 - 2 \approx 0.598076$.
These results show that the extended ensemble QMC simulation can be a vital tool to extract the boundary CFT entropies and surface free energies.

\subsection{Klein bottle entropy}

\begin{figure*}[!t]
  {\includegraphics[width=\textwidth]{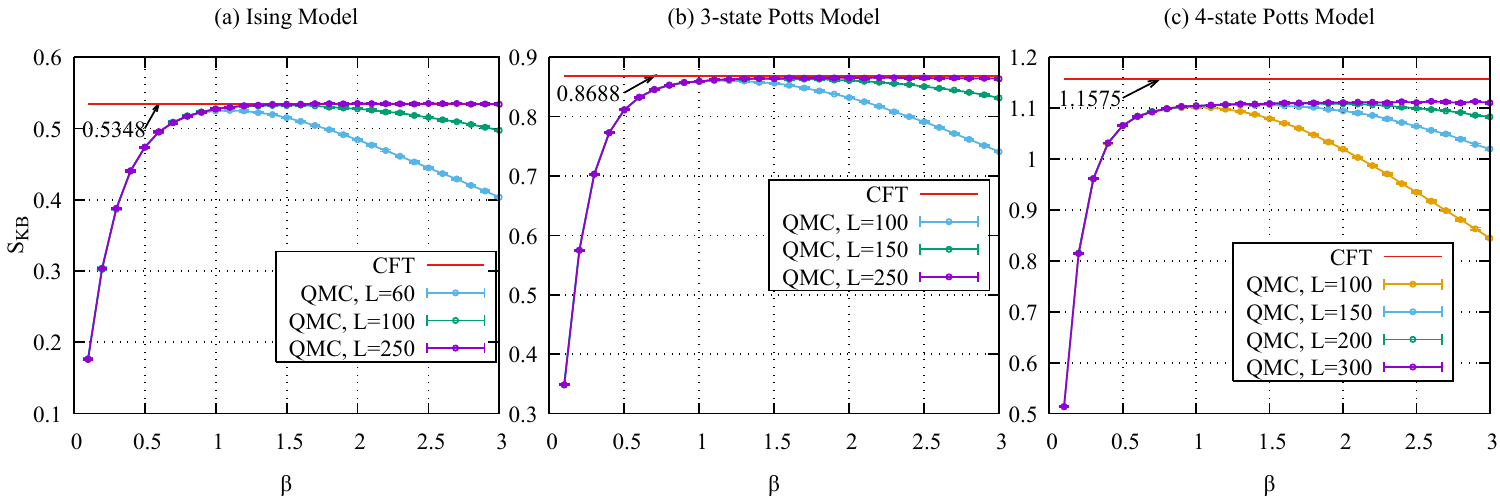}}
  \caption{Quantum Monte Carlo results of the Klein bottle entropies for (a) $q = 2$ (Ising) (b) $q =
  3$ (c) $q = 4$ quantum Potts chains. \label{fig:klein-result}}
\end{figure*}

The extraction of Klein bottle entropy is simpler than the AL entropy because there are no nonuniversal surface free energies in Eq.~(\ref{SKB}). The ratio should converge
to a constant with the increase of $\beta$, provided that $L \gg v \beta$ is also satisfied. Since the $q$-state quantum 
Potts chain (\ref{q-potts-hamiltonian}) exhibits a second order quantum phase transition only for $q \leq 4$ (while for $q
> 4$, it exhibits a
first order phase transition, which cannot be described by CFT), we calculate
the Klein bottle entropy for $q = 2, 3, 4$ and compare them with the CFT results. In the following, we first review the CFT predictions and then compare them with the numerical results.  

\paragraph{2-state quantum Potts (Ising) chain}When $q = 2$, the $q$-state quantum Potts model (\ref{q-potts-hamiltonian}) reduces to the
transverse field Ising model. The central charge of the Ising model is $c = 1 /
2$, so the critical point of Ising model is described by the free Majorana fermion CFT,
which has three primary fields{\cite{francesco_conformal_2012}}, $\mathbb{I}$ ($d_{\mathbb{I}} =
1$), $\psi$ ($d_{\psi} = 1$), and $\sigma$ ($d_{\sigma} = \sqrt{2}$). The
total quantum dimension $\mathcal{D} = 2$. Thus, the CFT prediction of Klein bottle entropy
for $q = 2$ Potts model is
\begin{equation}
  S_{\mathrm{KB}} = \mathrm{ln} \left( \frac{2 + \sqrt{2}}{2} \right) \approx
  0.5348.
\end{equation}

\paragraph{Three-state quantum Potts chain} At its quantum critical point, the low-energy effective theory of three-state quantum Potts chain is the $Z_3$ parafermion CFT with central charge $c=4/5$. Although a CFT with $c=4/5$ can be uniquely associated with the minimal model $\mathcal M(6,5)$ [the third unitary minimal model CFT $\mathcal M(p+1,p)$ with $p=5$, see Ref.~\onlinecite{francesco_conformal_2012}], there is a subtlety in understanding the CFT operator content for the three-state Potts chain: the minimal model $\mathcal M(6,5)$ has ten Virasoro primary fields, which are labeled by two integers $(n,m)$ with $1 \leq m \leq n \leq 4$. However, only six Virasoro primaries appear in the energy spectrum of the three-state Potts chain with periodic boundary conditions (see, e.g., Ref.~\onlinecite{li2015}). These six primaries are $(1,1)$, $(2,1)$, $(3,1)$, $(4,1)$, $(3,3)$, and $(4,3)$. The partition function  on a torus, which encodes the energy spectrum information, is the following nondiagonal modular invariant:
\begin{eqnarray}
Z^{\mathcal T} &=&|\chi_{1,1}(\mathfrak{q})+\chi_{4,1}(\mathfrak{q})|^2
+|\chi_{2,1}(\mathfrak{q})+\chi_{3,1}(\mathfrak{q})|^2 \notag \\
&\phantom{=}& +2|\chi_{3,3}(\mathfrak{q})|^2+2|\chi_{4,3}(\mathfrak{q})|^2 \, ,
\label{eq:Z3torus}
\end{eqnarray}%
where $\chi_{n,m}(\mathfrak{q})=\mathrm{Tr}_{(n,m)}(\mathfrak{q}^{L_{0}-c/24})$ is the Virasoro character with $\mathrm{Tr}_{(n,m)}$ being the trace within the Virasoro tower of the primary $(n,m)$. Here $\mathfrak{q}=e^{-2\pi{v\beta}/{L}}$ (not to be confused with the ``$q$'' of the $q$-state Potts model) and $L_0$ is the zero-th level Virasoro generator.

Following Ref.~\onlinecite{tu_characterizing_2017}, the Klein bottle partition function, for which only holomorphic-antiholomorphic symmetric
states in the conformal towers have contributions, is then given by
\begin{eqnarray}
Z^{\mathcal K} &=&\chi_{1,1}(\mathfrak{q}^2)+\chi_{4,1}(\mathfrak{q}^2)
+\chi_{2,1}(\mathfrak{q}^2)+\chi_{3,1}(\mathfrak{q}^2) \notag \\
&\phantom{=}& +2\chi_{3,3}(\mathfrak{q}^2)+2\chi_{4,3}(\mathfrak{q}^2) \, .
\label{eq:Z3klein}
\end{eqnarray}
When considering the limit $L\gg v\beta \; (\mathfrak{q} \rightarrow 1)$, one can use the modular transformation properties of the Virasoro characters to evaluate (\ref{eq:Z3klein}). This immediately leads to a universal entropy $S_\mathrm{KB}= \ln g$, where $g = (d_{1,1}+d_{2,1}+d_{3,1}+d_{4,1}+2d_{3,3}+2d_{4,3})/\mathcal{D}$. Here the quantum dimensions of the six Virasoro primaries are given by $d_{1,1} = d_{4,1} = 1$, $d_{2,1} = d_{3,1} = (1+\sqrt{5})/2$, $d_{3,3} = 1+\sqrt{5}$, and $d_{4,3} = 2$ (see Table 1 in Ref.~\onlinecite{affleck_boundary_1998}) and the total quantum dimension $\mathcal{D}$, which requires to take all ten primaries into account, is given by $\mathcal{D} = \sqrt{6(5+\sqrt{5})}$.
Thus, the Klein bottle entropy $S_{\mathrm{KB}}$ for the $Z_3$ parafermion CFT is given by
\begin{equation}
S_{\mathrm{KB}}  = \mathrm{ln} \left ( \sqrt{3+\frac{6}{\sqrt{5}}} \right)  \approx 0.8688 \, .
\label{eq:Z3entropy}
\end{equation}

For the three-state quantum Potts chain, it is also instructive to arrive at the above result for $S_{\mathrm{KB}}$ through an extended symmetry of the $Z_3$ parafermion CFT. By putting the Virasoro primary field $(4,1)$ into the chiral algebra, the Virasoro algebra is promoted to a $W$ algebra \cite{fateev1987}. At the partition function level, this corresponds to recombining the Virasoro characters into $W$ characters, $\chi_{\mathbb{I}} = \chi_{1,1}+\chi_{4,1}$, $\chi_\varepsilon = \chi_{2,1}+\chi_{3,1}$, $\chi_\sigma = \chi_{\sigma^\dag} = \chi_{3,3}$, and $\chi_\psi = \chi_{\psi^\dag} = \chi_{4,3}$, where the six primaries (with respect to the $W$ algebra) can now be denoted by a set $s= \{\mathbb{I}, \varepsilon, \sigma, \sigma^\dag, \psi, \psi^\dag \}$. In this framework, the torus partition function becomes diagonal, $Z^{\mathcal T} = \sum_{a \in s} |\chi_{a}(\mathfrak{q})|^2$, and the Klein bottle partition function is the sum of all six $W$ characters, $Z^{\mathcal K} = \sum_{a \in s} \chi_{a}(\mathfrak{q}^2)$. According to Ref.~\onlinecite{tu_characterizing_2017}, the ``ground-state degeneracy'' $g$ is then given by $g = \sum_{a \in s} d_a/\mathcal{D}'$, where the six primaries have quantum dimensions \cite{cardy_boundary_1989} $d_\mathbb{I} = d_{\psi} = d_{\psi^\dag} = 1$ and $d_{\varepsilon} = d_{\sigma} = d_{\sigma^\dag} = (1+\sqrt{5})/2$ and the total quantum dimension is given by $\mathcal{D}'=\sqrt{\sum_{a\in s}d^2_a}=\sqrt{\frac32(5+\sqrt{5})}$. It is easily verified that the value of $S_{\mathrm{KB}}$ so obtained is the same as (\ref{eq:Z3entropy}).

\paragraph{Four-state quantum Potts chain} The four-state quantum Potts chain is also critical at $\Gamma =J$ and its low-energy effective theory is the $Z_2$ orbifold of a $U(1)_8$ CFT with central charge $c=1$ \cite{ginsparg1988}. Alternatively, this CFT can be viewed as the $D_2 = Z_2 \times Z_2$ orbifold of the $SU(2)_1$ CFT \cite{dijkgraaf1989,barkeshli2014}. Here we follow the notation of Ref.~\onlinecite{dijkgraaf1989} and denote the set of eleven primaries as $t= \{ \mathbb{I}, j_b, \phi, \sigma_b, \tau_b \}$, where $b=1,2,3$. In this case, the torus and Klein bottle partition functions are the usual diagonal modular invariant and a sum of all eleven chiral characters, respectively. Then, we have $g=\sum_{a\in t}d_a/\mathcal{D}$. Here the quantum dimensions of the primaries are $d_{\mathbb{I}}=d_{j_b} = 1$ and $d_{\phi}=d_{\sigma_b} = d_{\tau_b}=2$ (see Table 6 in Ref.~\onlinecite{dijkgraaf1989}), and the total quantum dimension is $\mathcal{D} = \sqrt{\sum_{a\in t}d^2_a}=4\sqrt{2}$. Then, we obtain the value of Klein bottle entropy
\begin{equation}
S_\mathrm{KB} = \mathrm{ln}\left( \frac{9}{2\sqrt{2}} \right ) \approx 1.1575 \, .
\label{eq:Z4entropy}
\end{equation}

\paragraph{Simulation results and comparison with CFT predictions}The simulation results of $q = 2, 3$ are shown in Fig.~\ref{fig:klein-result}(a,b), where $S_{\mathrm{KB}}$ is estimated by
$\mathrm{ln} \left[ Z^{\mathcal{K}} (2 L, \beta / 2) / Z^{\mathcal{T}} (L, \beta) \right]$
according to Eq.~(\ref{SKB}). As a comparison, the CFT predictions are
marked as a horizontal line correspondingly. As can be seen from the figure, for
small $\beta$, the QMC estimations strongly deviate from the CFT predictions due to
the existence of the correlation term $\mathcal{O} (1 / \beta^2)$ in \Eq{SKB}. While as $\beta$ increases the QMC results reach a 
plateau value, which is in agreement with the CFT predictions of
$S_{\ensuremath{\operatorname{KB}}}$. The plateau is more visible for larger system sizes because the
condition $L \gg v\beta$ is better satisfied in the simulated temperature range. Overall, we see excellent agreement of the CFT predictions of the Klein bottle entropy with the QMC results for $q=2,3$ quantum Potts chain. While for the four-state quantum Potts chain, the boundary entropy saturates into a value that is slightly smaller than the CFT prediction [see Fig.~\ref{fig:klein-result}(c)]. We conjecture that the deviation is attributed to a marginally irrelevant term~\cite{nauenberg_singularities_1980, cardy_scaling_1980, blote_scaling_2017} in the low-energy field theory, which is known to be present in the four-state Potts model. We performed exact diagonalization of the four-state quantum Potts chain with moderate size and found that the eigenenergies of several low-energy states indeed deviate from the CFT predictions, which may be seen as the effect of marginally irrelevant perturbations. The effect of marginally irrelevant terms on the Klein bottle entropy is beyond the scope of the current work and will be reported elsewhere~\cite{WeiLiinprep}.


\section{Summary}\label{conclusion}

To summarize, we have presented an efficient Monte Carlo algorithm to extract
the universal boundary entropies from the lattice models. The simulation techniques developed in this paper not only provide a way to extract boundary entropies of quantum lattice models in a general setting, but also provides a new and unified picture of Affleck-Ludwig and Klein entropies of conformal field theories: the Klein bottle entropy
can be interpreted as a boundary entropy through a transformation of the path-integral
manifold.

From the numerical side, we have computed Affleck-Ludwig boundary
entropies in $q = 2, 3$ quantum Potts chains with various boundary conditions, as well as their Klein bottle entropies, which show excellent agreement with the CFT predictions. We have also observed that the Klein entropy for the $q=4$ Potts model is slightly smaller than the CFT prediction, which we attribute to marginally irrelevant interactions in the low energy effective theory. A detailed study of the Klein bottle entropy under renormalization group flows may establish similar results such as the $g$-theorem~\cite{friedan_boundary_2004} for the Affleck-Ludwig boundary entropy.

\section*{Acknowledgement}

We are grateful to Meng Cheng for stimulating discussions.
This work is supported by NSF-China under Grant No.11504008 (W.T. and X.C.X) and Ministry of Science and Technology of China under the Grant No.2016YFA0302400 (L.W.). HHT acknowledges the support from the DFG through the Excellence Cluster ``Nanosystems Initiative Munich''.
The simulation is performed at Tianhe-1A platform at the National Supercomputer Center in Tianjin.

\clearpage
\appendix

\section{Continuous-time path integral Monte Carlo algorithm for \boldmath{$q$}-state
quantum Potts model}\label{app-update-scheme}

\paragraph{General description} The $q$-state quantum Potts chain Hamiltonian (\ref{q-potts-hamiltonian}) can be
splitted into its diagonal part $\hat{H}_0$ and its off-diagonal part $\hat{H}_1$:
\begin{eqnarray}
  \hat{H}_0 & = & - J \sum_{\langle i, j\rangle}  \sum_{k = 1}^{q - 1} \hat{\sigma}_i^k
   \hat{\sigma}_{j}^{q - k},  \label{qpottsh2}\\
  \hat{H}_1 & = & - \Gamma \sum_{i = 1}^L \sum_{k = 1}^{q - 1} \hat{\tau}_i^k.
  \label{qpottsh1}
\end{eqnarray}
For the diagonal part $\hat{H}_0$, the matrix element is
\begin{equation}
  \langle \ensuremath{\boldsymbol{\sigma}} | \hat{H}_0 |
  \ensuremath{\boldsymbol{\sigma}} \rangle
  = \sum_{\langle i, j \rangle} \left( J - \delta_{\sigma_i, \sigma_j} q J \right)
  \equiv \sum_{\langle i, j \rangle} \mathcal{J}
  (\sigma_i, \sigma_j), \label{h0-mat-elem}
\end{equation}
where $\ensuremath{\boldsymbol{\sigma}} = (\sigma_1, \sigma_2, \ldots,
\sigma_L)$, $| \ensuremath{\boldsymbol{\sigma}} \rangle \equiv | \sigma_1
\rangle \otimes | \sigma_2 \rangle \otimes \ldots \otimes | \sigma_L \rangle$,
and $\langle i, j \rangle$ represents the two neighboring sites $i, j$.
While for the transverse field $\hat{H}_1$, one can obtain that
\begin{equation}
  \langle \ensuremath{\boldsymbol{\sigma}} | \hat{H}_1 |
  \ensuremath{\boldsymbol{\sigma}}' \rangle =
  \sum_{i} \left( -\Gamma (1 - \delta_{\sigma_{i},\sigma_{i}'}) \prod_{j \neq i} \delta_{\sigma_{j},\sigma_{j}'} \right),
  \label{h1-mat-elem}
\end{equation}
which implies that in order to make the matrix element $\langle
\ensuremath{\boldsymbol{\sigma}} | \hat{H}_1 |
\ensuremath{\boldsymbol{\sigma}}' \rangle$ nonzero,
$\ensuremath{\boldsymbol{\sigma}}$ and $\ensuremath{\boldsymbol{\sigma}}'$
should differ with each other by only one spin at site $i$, which is summed over in the out summation.

Write $e^{- \beta \hat{H}}$ in the interaction representation
\begin{equation}
  e^{- \beta \hat{H}} = e^{- \beta \hat{H}_0}
  \mathrm{T}_{\tau} \left[ \exp \left( - \int_0^{\beta} \hat{H}_1 (\tau)
  \mathrm{d} \tau \right) \right],
\end{equation}
where $\hat{H}_1 (\tau) = e^{\tau \hat{H}_0} \hat{H}_1 e^{-
\tau \hat{H}_0}$, $\mathrm{T}_{\tau}$ is the time-ordering operator. Expanding the exponential in Taylor series, and rearranging the terms to cancel the $k!$ in the denominator, the torus partition function reads 

\begin{widetext}
\begin{equation}
Z^{\mathcal{T}}(L, \beta) =\mathrm{Tr} \left[e^{- \beta  \hat{H}_0} \sum_{k = 0}^{\infty} (- 1)^k
 \int_0^{\beta} \mathrm{d} \tau_1 \int_{\tau_1}^{\beta} \mathrm{d} \tau_2 \ldots \int_{\tau_{k -1}}^{\beta} \mathrm{d} \tau_k
 \hat{H}_1 (\tau_1) \hat{H}_1 (\tau_2) \ldots \hat{H}_1 (\tau_k) \right] ,
\end{equation}
We write the trace explicitly by
$\sum_{\ensuremath{\boldsymbol{\sigma}}} \langle \ensuremath{\boldsymbol{\sigma}} |\cdot| \ensuremath{\boldsymbol{\sigma}} \rangle$,
and insert the resolution of the identity $1 =
\sum_{\ensuremath{\boldsymbol{\sigma}}} | \ensuremath{\boldsymbol{\sigma}}
\rangle \langle \ensuremath{\boldsymbol{\sigma}} |$ around each $\hat{H}_1$.
Using Eq.~(\ref{h1-mat-elem}), one can obtain
\begin{eqnarray}
  Z^{\mathcal{T}} (L, \beta) & = & \sum_{k = 0}^{\infty} \sum_{\ensuremath{\boldsymbol{\sigma}}^{(0)}
  \ldots \ensuremath{\boldsymbol{\sigma}}^{(k)}}
  \int_0^{\beta} \mathrm{d} \tau_1 \int_{\tau_1}^{\beta} \mathrm{d} \tau_2 \ldots \int_{\tau_{k -1}}^{\beta} \mathrm{d} \tau_k \,
  \Gamma^k \prod_{m = 0}^k \langle \ensuremath{\boldsymbol{\sigma}}^{(m)} | e^{-
  (\tau_m - \tau_{m + 1}) \hat{H}_0} | \ensuremath{\boldsymbol{\sigma}}^{(m+1)}
  \rangle  \label{ctqmc2}\\
  & = & \sum_{k = 0}^{\infty} \sum_{\ensuremath{\boldsymbol{\sigma}}^{(0)}
  \ldots \ensuremath{\boldsymbol{\sigma}}^{(k)}}
  \int_0^{\beta} \mathrm{d} \tau_1 \int_{\tau_1}^{\beta} \mathrm{d} \tau_2 \ldots \int_{\tau_{k -1}}^{\beta} \mathrm{d} \tau_k \,
  \Gamma^k \exp \left( - \sum_{m = 0}^k \sum_{\langle i, j \rangle} (\tau_m -
  \tau_{m + 1}) \mathcal{J} (\sigma_i^{(m)}, \sigma^{(m)}_j) \right) . \label{Z-continuous}
\end{eqnarray}
\end{widetext}
where $0 = \tau_0 \leq \tau_1 \leq \ldots \leq \tau_{k+1} = \beta$ and $\ensuremath{\boldsymbol{\sigma}}^{(k+1)}\equiv \ensuremath{\boldsymbol{\sigma}}^{(0)}$. We then can sample the partition function (\ref{Z-continuous}) by Monte Carlo
method. From Eq. (\ref{Z-continuous}), the Boltzmann weight of
configuration $\mathscr{C}=\{ \tau_{m}, \ensuremath{\boldsymbol{\sigma}}^{(m)} \}$ can be written as
\begin{equation}
 w^{\mathcal{T}} (\mathscr{C}) = \Gamma^k \exp \left( - \sum_{m = 0}^k \sum_{\langle i, j
  \rangle} (\tau_m - \tau_{m + 1}) \mathcal{J} (\sigma_i^{(m)}, \sigma^{(m)}_j) \right) .
  \label{boltzmann-weight}
\end{equation}
Notice that according to Eq~(\ref{h1-mat-elem}), $\ensuremath{\boldsymbol{\sigma}}^{(m)}$ and $\ensuremath{\boldsymbol{\sigma}}^{(m+1)}$ differ with each other by one spin flip. The configurations can now be represented by a set of worldlines with vertices on them, and each of these vertices represents an off-diagonal term in 
$\hat{H}_1$ at the imaginary time $\{\tau_m\}$. It is worth noting that since $\Gamma> 0$, \Eq{boltzmann-weight} is positive definite for any valid configuration and can be directly interpreted as a probability density.

While for the partition with conformal boundary conditions $Z^{\mathcal{C}}(L, \beta)$ can be computed by modifying the boundary conditions in the Hamiltonian. We discuss implementation of these conformal boundary conditions in Appendix \ref{appendix:complete-set-of-bc}. Moreover, to sample $Z^{\mathcal{K}}(2L, \beta/2)$, we transform the Klein bottle to a cylinder with size $L$ and temperature $\beta$ with nonlocal interactions along imaginary time direction on the boundary sites, as discussed in Sec.~\ref{ratio}.
%


\paragraph{Cluster update}Due to the equivalence between the (1+1)-dimensional
quantum model and a two-dimensional classical model{\cite{suzuki_relationship_1976}},
we can promote the Swendsen-Wang cluster algorithm{\cite{swendsen_nonuniversal_1987}}
in the classical cases to the continuous time limit{\cite{rieger_application_1999}}. The cluster update algorithm identifies clusters with the same states and changes the worldline configurations collectively and randomly. 

To map the path-integral configuration of a continuous-time QMC simulation of 1D quantum system to a 2D classical lattice model, one can divide the imaginary-time axis
into many small segments of size $\Delta \tau$,
\begin{widetext}
\begin{equation}
  Z^{\mathcal{T}} = \sum_{\{ \ensuremath{\boldsymbol{\sigma}}^{(l)} \}}
  \exp \left[ \sum_l^{\beta / \Delta \tau} \left(\sum_{i}
  (1 - \delta_{\sigma_i^{(l)}, \sigma_i^{(l+1)} } ) \mathrm{ln}(\Delta \tau \Gamma)
  - \sum_{\langle i,j \rangle} \Delta \tau \mathcal{J} (\sigma_{i}^{(l)}, \sigma_{j}^{(l)}) \right)\right] \equiv  \sum_{\{ \ensuremath{\boldsymbol{\sigma}}^{(l)} \}}
  \exp \left[ - H_{\mathrm{eff}} \right], \label{effective-Hamiltonian}
\end{equation}
\end{widetext}
where $l$ is the index of the segments along the imaginary time direction. We obtain an effective classical Hamiltonian $H_{\mathrm{eff}}$, which reads up to a constant term
\begin{equation}
  H_{\mathrm{eff}} = \sum_l^{\beta / \Delta \tau} \left( K^{\tau}  \sum_{i } \delta_{\sigma_{i}^{(l)}, \sigma_{i}^{(l+1)} } +  K^{x} \sum_{\langle{i,j}\rangle}  \delta_{\sigma_{i}^{(l)}, \sigma_{j}^{(l)}}\right). \label{q-state-eff}
\end{equation}
So the quantum Potts chain is equivalent to the two-dimensional
classical Potts model{\cite{wu_potts_1982}} with anisotropic
coupling parameters $K^{\tau} = \ln (\Delta \tau \Gamma), K^x = - q J \Delta \tau$ in the two dimensions. Provided $\Delta \tau$ is small enough, both couplings are ferromagnetic (smaller than zero).

Now one can apply the Swendsen-Wang algorithm to the classical model (\ref{q-state-eff}).
The probability of connecting two sites with the same states along the imaginary time direction is
\begin{equation}
  P_{\text{add}}^{\tau} = 1 - \Gamma \Delta \tau . \label{add-along-tau}
\end{equation}
The probability of connecting two sites with the same state along the spatial direction is
\begin{equation}
  P_{\text{add}}^x = 1 - \exp (\Delta \tau q J) \approx \Delta \tau q J.
  \label{add-along-x}
\end{equation}
Back to the continuous-time configuration ($\Delta \tau \rightarrow 0$), the process of generating clusters is replaced by generating clusters by adding cuts and connections in
the configuration{\cite{rieger_application_1999}}. It can be seen from
equation (\ref{add-along-tau}, \ref{add-along-x}), in our case, we can add
cuts along the time line by Poisson process with density $\Gamma$, and add
bonds between the time lines to connect the segments of the same state by Poisson process with density $q J$.

If an external magnetic field is present (in our case, the boundary pinning field), the effect of the magnetic field can be integrated into the process of the flipping of the clusters. To be more specific, one can follow the ``two-step selection'' procedure\cite{gubernatis_quantum_2016}, and write the overall transition matrix as
\begin{equation}
  P(\mathscr{C}', \mathscr{C})=\sum_{\mathscr{G}} P(\mathscr{C}', \mathscr{G}) P(\mathscr{G},\mathscr{C}) ,
\end{equation}
where $\mathscr{C},\mathscr{C}'$ are the different configurations, $\mathscr{G}$ is the graph produced by the bond generation procedure, $P(\mathscr{G},\mathscr{C})$ is the probability to produce the graph $\mathscr{G}$ from the configuration $\mathscr{C}$, and $P(\mathscr{C}', \mathscr{G})$ is the probability to generate the configuration $\mathscr{C}'$ from the graph $\mathscr{G}$. Then the detailed balance condition can be satisfied if the following equation holds for any graph $\mathscr{G}$
\begin{equation}
  P(\mathscr{C}', \mathscr{G}) P(\mathscr{G},\mathscr{C}) w(\mathscr{C}) = P(\mathscr{C}, \mathscr{G}) P(\mathscr{G},\mathscr{C}) w(\mathscr{C}').
  \label{eq:detailed-balance-two-step}
\end{equation}
If the external magnetic field is absent, it has been proven\cite{gubernatis_quantum_2016} that this equation is satisfied by the normal Swendsen-Wang procedure with $P(\mathscr{C}', \mathscr{G}) = P(\mathscr{C}, \mathscr{G}) = 1/q^{N_{\mathrm{clusters}}}$ (each possible state is assigned to a cluster with even probability $1/q$). On the other hand, if there is an external magnetic field, we can split the Boltzmann weight of the configuration as $w(\mathscr{C}) = w_{0}(\mathscr{C}) w_{\mathrm{ext}}(\mathscr{C})$, where $w_{\mathrm{ext}}(\mathscr{C})$ is the contribution of the external-field term to the Boltzmann weight. One can see that Eq.~(\ref{eq:detailed-balance-two-step}) can be written as
\begin{eqnarray}
  & & P(\mathscr{C}', \mathscr{G}) P(\mathscr{G}, \mathscr{C}) w_{0}(\mathscr{C}) w_{\mathrm{ext}}(\mathscr{C}) \nonumber \\
  & = & P(\mathscr{C}, \mathscr{G}) P(\mathscr{G},\mathscr{C}') w_{0}(\mathscr{C}') w_{\mathrm{ext}}(\mathscr{C}')
\end{eqnarray}
From the discussion above one can infer that $P(\mathscr{G}, \mathscr{C}) w_{0}(\mathscr{C}) = P(\mathscr{G},\mathscr{C}') w_{0}(\mathscr{C}')$, so
\begin{equation}
  P(\mathscr{C}', \mathscr{G}) w_{\mathrm{ext}}(\mathscr{C}) = P(\mathscr{C}, \mathscr{G}) w_{\mathrm{ext}}(\mathscr{C}'), \forall \mathscr{G}.
  \label{eq:detailed-balance-external-field}
\end{equation}
Eq.~(\ref{eq:detailed-balance-external-field}) can be satisfied by flipping each cluster in the following way. For a cluster whose spins are originally in the state $A$, the probability of setting this cluster to state $A'$ is determined by
\begin{align}
  p_{A'}e^{-\beta E_{\mathrm{ext}}(A)} &= p_{A} e^{-\beta E_{\mathrm{ext}}( A')}, \label{eq:detailed-balance-cluster} \\
  \sum_{A} p_{A} &= 1. \label{eq:sum-probability}
\end{align}
Here $p_{A}$ is the probability of setting the cluster to state $A$, and $E_{\mathrm{ext}}(A)$ is the coupling energy caused by the external magnetic field of the cluster in the state $A$. Eq.~(\ref{eq:detailed-balance-cluster})(\ref{eq:sum-probability}) can be satisfied by choosing
\begin{equation}
  p_{A} = \frac{e^{-\beta E_{\mathrm{ext}}(A)}}{\sum_{A'} e^{-\beta E_{\mathrm{ext}}( A')}}
\end{equation}

\section{Conformal boundary conditions of Ising model and three-state Potts
model}\label{appendix:complete-set-of-bc}

A complete set of boundary states (and hence boundary conditions) of Ising
model and three-state Potts model can be generated by CFT. For the Ising model, the
complete set of boundary conditions include free and fixed boundary
condition{\cite{cardy_boundary_1989}}. For the three-state Potts model, the complete
set of boundary conditions include free, fixed, mixed and
symmetric\footnote{In the literature, the symmetric boundary condition is often referred to as the ``new'' boundary condition.
We call it ``symmetric'' since this boundary condition favours the $Z_3$ symmetric state.}
boundary condition{\cite{cardy_boundary_1989,saleur_relations_1989,affleck_boundary_1998,fuchs_completeness_1998}}.
In this appendix, we summarize the CFT results of the complete set of boundary conditions
for the Ising and three-state Potts model. We also discuss ways to achieve various boundary conditions in the quantum spin chain.

\paragraph{Ising Model}The Hamiltonian of quantum Ising chain with \emph{free}
boundary condition is
\begin{equation}
  \hat{H}_\mathrm{free} = - J \sum_{i = 1}^{L - 1} \hat{\sigma}_i \hat{\sigma}_{i + 1} -
  \Gamma \sum_{i = 1}^L \hat{\tau}_i .
\end{equation}
To achieve the \emph{fixed} boundary condition, we can add a very strong longitudinal pinning field on the boundary sites:
\begin{equation}
  \hat{H}_\mathrm{fixed} = \hat{H}_\mathrm{free} - h_{\mathrm{b}} (\hat{\sigma}_1 +
  \hat{\sigma}_L).
\end{equation}
The simulation remains sign problem free.

\paragraph{three-state Potts Model}The Hamiltonian of the  three-state Potts chain with free boundary condition
is
\begin{equation}
  \hat{H}_\mathrm{free} = - J \sum_{j = 1}^{L - 1} \left( \hat{\sigma}_j^2
  \hat{\sigma}_{j + 1} + \text{h.c.} \right) - \Gamma \sum_{j = 1}^L
  (\hat{\tau}_j + \hat{\tau}_j^2) .
\end{equation}
The three possible states of each site will be denoted as $|\mathrm{A} \rangle, |\mathrm{B}
\rangle, |\mathrm{C} \rangle$. As described above, the fixed boundary condition
``favors'' one of the three states of the Potts model, while the mixed
boundary condition ``forbids'' one of the three states. Similar to the case
of the Ising model, the fixed and mixed boundary condition can be achieved by
adding a complex valued longitudinal magnetic field to the free boundary
Hamiltonian,
\begin{eqnarray}
  \hat{H}_\mathrm{fixed/mixed} =    \hat{H}_\mathrm{free} &-& \left( h_{\mathrm{b}} \hat{\sigma}_1 + h_{\mathrm{b}}^{\ast}
  \hat{\sigma}_1^2 \right)  \nonumber\\
  & - &
   \left( h_{\mathrm{b}} \hat{\sigma}_L +
  h_{\mathrm{b}}^{\ast} \hat{\sigma}_L^2 \right) .
\end{eqnarray}
The longitudinal field $h_\mathrm{b}$ appears  in the exponent of \Eq{Z-continuous} and there is no sign problem for any value of $h_\mathrm{b}$. As an example, assume now that we choose to favor or forbid $|\mathrm{A} \rangle$, we can set $h_{\mathrm{b}}$ to be real (i.e. the complex phase of $h_{\mathrm{b}}$ is 0), then the boundary term becomes
\begin{equation}
  - \left( h_{\mathrm{b}} \hat{\sigma} + h_{\mathrm{b}}^{\ast}
  \hat{\sigma}^2 \right) = h_{\mathrm{b}} \left(
  \begin{array}{ccc}
    - 2 &  & \\
    & 1 & \\
    &  & 1
  \end{array} \right) .
\end{equation}

When $ h_{\mathrm{b}} \rightarrow +\infty$, then both boundary sites of the Potts chain are pinned to Potts state $|\mathrm{A} \rangle$ and we then achieve the fixed boundary condition. On the other hand, to obtain the mixed boundary condition, we can set $h_{\mathrm{b}} \rightarrow -\infty $ to forbid the state $|\mathrm{A} \rangle$. Similarly, to favor or forbid other Potts states $|\mathrm{B} \rangle$ and $|\mathrm{C} \rangle$, one only needs to modify the complex phase of the pinning field to be $2\pi/3$ or $4\pi/3$.

There is another conformal boundary condition{\cite{affleck_boundary_1998}} in addition to the free, fixed and mixed boundary
condition for the three-state Potts model.
At the critical point (where $\Gamma = J$), the symmetric boundary
condition can be achieved by adding a complex transverse field at the boundary
sites to the free boundary condition
\begin{eqnarray}
  \hat{H}_\mathrm{sym} =   \hat{H}_\mathrm{free}  &-& (h_\mathrm{T} \hat{\tau}_1 + h_\mathrm{T}^{\ast} \hat{\tau}_1^2) \nonumber\\
  & - &  (h_\mathrm{T}
  \hat{\tau}_L + h_\mathrm{T}^{\ast} \hat{\tau}_L^2) .
\end{eqnarray}
When $h_\mathrm{T} = - J$  the transverse pinning field favors a $Z_3$ symmetric state therefore achieving the
symmetric boundary condition. However, this Hamiltonian will encounter a sign problem in the
Monte Carlo simulation because of the negative transverse field on the boundary. To obtain the partition function of the three-state Potts
chain under the symmetric boundary condition, we use the following relation
obtained by means of Kramers-Wannier duality
transformation{\cite{affleck_boundary_1998}}
\begin{equation}
  Z^{\mathcal{C}}_{\mathrm{sym}, \text{sym}} = Z^\mathcal{C}_{\mathrm{AB,AB}} + Z^\mathcal{C}_{\mathrm{AB,AC}} +
  Z^\mathcal{C}_{\mathrm{AB,BC}},
  \label{eq:Zsymm}
\end{equation}
where the double subscripts of the partition function correspondingly
represent the boundary conditions on the two boundaries of the system.
$Z_{\text{sym}, \text{sym}}$ represents the partition function with symmetric
boundary conditions on both boundaries. $Z_{\mathrm{XY}, \mathrm{X}'
\mathrm{Y}'} \left( \mathrm{X}, \mathrm{Y}, \mathrm{X}', \mathrm{Y}' =
\mathrm{A}, \mathrm{B}, \mathrm{C} \right)$ represents the partition function
with mixed boundary conditions, and on the left boundary the states $|
\mathrm{X} \rangle, | \mathrm{Y} \rangle$ are degenerate, while on the right
boundary, the states $| \mathrm{X}' \rangle, | \mathrm{Y}' \rangle$ are
degenerate. So the partition function $Z_{\text{sym}, \text{sym}}$ can be
obtained by performing three independent Monte Carlo simulations to compute
the three terms in \Eq{eq:Zsymm} separately.

\section{Reweighting in the presence of boundary pinning field}\label{appendix-reweighting}

It requires special attention to construct the extended ensemble for systems involves strong pinning field because in these cases $Z^{\mathcal{C}} \gg  Z^{\mathcal{T}} $.
By Monte Carlo reweighting we simulate an extended ensemble with the reweighed summation of partition functions instead
of \Eq{eq:EE} in the main text.
For quantum Ising ($q=2$) chain
\begin{equation}
  Z = Z^{\mathcal{C}} +   e^{\beta \left( 2
h_{\mathrm{b}} - J \right)} Z^{\mathcal{T}} .
\end{equation}
The estimator \Eq{eq:BAR} in the extended ensemble simulation will now give
$e^{-\beta (2 h_{\mathrm{b}} - J)}(Z^{\mathcal{C}}/Z^{\mathcal{T}})$.
The prefactor will not affect the result of AL entropy $\ln (g)$, since it can be absorbed
into the nonuniversal surface energy term.

The prefactor before the second term ensures that the transitions between the configuration spaces are
possible to be accepted in both directions.
The energy difference of the transition from the torus configuration space to the cylinder configuration space
now becomes
\begin{equation}
  \Delta E_{\left\{ \mathcal{T} \rightarrow \mathcal{C} \right\}} = J
  (\hat{\sigma}_1^z \hat{\sigma}_L^z - 1) - h_{\mathrm{b}} (\hat{\sigma}_1^z + \hat{\sigma}_L^z - 2) .
\end{equation}
The energy difference $\Delta E_{\left\{ \mathcal{T} \rightarrow
\mathcal{C} \right\}}$ becomes 0 if the boundary spins are fully polarized, in which case
the transition is bound to be accepted in both transition directions.
The reweighting increases the acceptance rate and decreases the correlation time of the Monte Carlo simulation, and
thus yields a higher efficiency of the simulation.

In the case of the three-state Potts model, the Monte Carlo reweighting can be
performed in a similar way. In the case of fixed boundary
condition ($h_{\mathrm{b}} > 0$, $J > 0$), we multiply $Z^{\mathcal{T}}$ by
$e^{\beta \left( 4 h_{\mathrm{b}} - 2 J \right)}$, while in the case
of mixed boundary condition ($h_{\mathrm{b}} < 0$, $J > 0$), we multiply
$Z^{\mathcal{T}}$ by ${e}^{\beta \left( - 2 h_{\mathrm{b}} - 2 J
\right)}$\footnote{For the mixed boundary condition, the additional
coefficient ${e}^{- 2 \beta h_{\mathrm{b}}}$ cannot make the transition
bound to be accepted when the boundary spins satisfy the corresponding mixed
boundary condition. But it still helps improve the efficiency of the
simulation.}. 

\bibliography{ctqmc}

\end{document}